\documentclass[aps,showpacs,twocolumn]{revtex4}
\usepackage{epsfig}
\usepackage{amsmath}
\begin{document}

\def\ket#1{|#1\rangle}
\def\bra#1{\langle#1|}
\def\scal#1#2{\langle#1|#2\rangle}
\def\matr#1#2#3{\langle#1|#2|#3\rangle}
\def\keti#1{|#1)}
\def\brai#1{(#1|}
\def\scali#1#2{(#1|#2)}
\def\matri#1#2#3{(#1|#2|#3)}
\def\bino#1#2{\left(\begin{array}{c}#1\\#2\end{array}\right)}
\def\Ninf{$N\!\to\!\infty$}
\def\nm1{n\!-\!1}

\title{Coulomb analogy for nonhermitian degeneracies near quantum phase transitions}
\author{Pavel Cejnar$^{1,2}$}
\author{Stefan Heinze$^{3}$}
\author{Michal Macek$^{1}$}
\affiliation{
$^1$Institute of Particle and Nuclear Physics, Faculty of Mathematics and Physics, Charles University,
V Hole\v sovi\v ck\'ach 2, 180\,00 Prague, Czech Republic\\
$^2$European Center for Theoretical Studies in Nuclear Physics and Related Areas, 38050 Vilazzano (Trent), Italy\\
$^3$Institute of Nuclear Physics, University of Cologne, Z\"ulpicherstrasse 77, 50937 Cologne, Germany
}
\date{\today}

\begin{abstract}
Degeneracies near the real axis in a complex-extended parameter space of a hermitian Hamiltonian are studied.
We present a method to measure distributions of such degeneracies on the Riemann sheet of a selected level and apply it in classification of quantum phase transitions.
The degeneracies are shown to behave similarly as complex zeros of a partition function.
\end{abstract}

\maketitle

Quantum phase transitions (QPTs) appear in {\em infinite\/} lattice systems (like spin arrays with bound range of interactions) \cite{Voj03} as well as in {\em finite\/} many-body systems of interacting bosons or fermions, see e.g. Refs.\cite{Gil79,Die80,Zha87,Hei88,Row98,Ema03,Iac04,Cej05,Dus05,Ley05,Cej06,Cap07}.
QPTs are driven by interaction parameters of the Hamiltonian and may affect the ground state \cite{Gil79,Die80,Zha87,Hei88,Row98,Ema03,Iac04,Cej05,Dus05} as well as individual excited states \cite{Ley05,Cej06,Cap07}.
Although the QPT nonanalytic features strictly occur only in the system's infinite size limit, significant precursors can be observed already at a moderate size.

Typical examples of finite many-body systems with QPTs are provided by two-level bosonic models, where one of the bosons, $s$, is scalar, while the other, $b^{(L)}$, transforms under rotations as the $L^{\rm th}$rank tensor \cite{Dus05}.
The single-particle Hilbert space has a dimension $2L\!+\!2$, hence the term \lq\lq finite\rq\rq.
Such models with $L=0,1,2$ are used to describe various types of collective excitations in atomic nuclei and molecules \cite{Iac87,Iac95}.
In absence of interactions between $s$ and $b$ bosons, the ground state is a pure one-component condensate, usually the one with $s$ bosons (the system is in its \lq\lq spherical phase\rq\rq).
As the interaction strength $\xi$ increases, a certain critical point $\xi_{\rm c}$ is reached where the ground state becomes a mixture of both types of bosons (the \lq\lq deformed phase\rq\rq\ arises).

An important distinction between the infinite and finite models follows from the fact that in the latter ones a value $\propto N^{-1}$, where $N$ is the number of particles, can be identified with the Planck constant.
Hence \Ninf\ is the classical limit \cite{Gil79}.
Although this implies that QPTs in finite many-body models are rooted in semiclassical properties, there exists a plethora of genuinely quantum signatures.
Some of them will be discussed in this Letter, where a link is elaborated between the occurrence of various types of QPTs and the distribution of degeneracies of Hamiltonian eigenvalues in the complex-extended parameter space.

We will consider a class of Hamiltonians depending linearly on a single dimensionless control parameter $\lambda$,
\begin{equation}
H(\lambda)=H_0+\lambda V
\,,
\label{ham}
\end{equation}
where $[H_0,V]\!\neq\!0$.
The Hilbert space dimension $n$ is finite and energy eigenvalues $E_k(\lambda)$ are counted from $k$=0 (the ground state) to $k\!=\!\nm1$.
As generally known, the maximal rate of change of a given eigenvector $\ket{\psi_k(\lambda)}$ appears when the level $k$ undergoes an avoided crossing.
A QPT takes place if the relative spacing between levels (in units of the mean spacing) becomes infinitely small as \Ninf.
If the limiting process results in a crossing of two levels, the QPT is of the first order (with a jump of $\tfrac{d}{d\lambda}E_k$ and the corresponding swap of wave functions).
On the other hand, if the number of interacting levels is locally large but with no real crossing, the resulting QPT is continuous (with a softer type of nonanalyticity).

The occurrence of avoided crossings in the spectrum signals the presence of true degeneracies somewhere nearby, in the plane of complex-extended parameter $\Lambda\!\equiv\!\lambda\!+\!i\mu$ \cite{Hei88,Rot01}.
The generalized Hamiltonian $H(\Lambda)$ is non-hermitian and requires to handle the left and right eigenvectors.
The eigenvalues live on $n$ Riemann sheets which can be labeled by the ordinal number $k$ of the respective level on the real axis.
There can be two simplest types of degeneracies: (i) a diabolic point, when the two sheets just touch each other \cite{Ber84}, and (ii) an exceptional, or branch point \cite{Kat66}, if the two sheets are entangled by the square-root type of singularity \cite{Zir83,Hei91,Gun07}.
In our case, the common type of degeneracy is (ii).
This follows from a perturbative expansion near (but not at) the degeneracy which allows to approximate the evolution of the two close eigenvalues by a 2$\times$2 matrix with generically nonzero offdiagonal elements.
In this situation, the degeneracy is a branch point.
A diabolic point would require an additional constraint that offdiagonal matrix elements vanish at the singularity \cite{Gun07}.
Since in the following we do not check individual nature of each crossing, we use a general term \lq\lq nonhermitian degeneracy\rq\rq\ \cite{Ber04}.

If a nonhermitian degeneracy comes close to the real $\Lambda$ axis, an avoided crossing is encountered in level dynamics of $H(\lambda)$.
We will therefore focus on the distribution of degeneracies in the $\Lambda$ plane.
The degeneracies are simultaneous roots of the characteristic polynomial $P(E,\Lambda)\equiv{\rm det}[E-H(\Lambda)]$ and its derivative $\tfrac{\partial}{\partial E}P(E,\Lambda)$.
The elimination of variable $E$ leads to the discriminant $D(\Lambda)$ (proportional to the resultant), a polynomial of order $n(\nm1)$, whose roots indicate positions $\Lambda_{\alpha}$ of the degeneracies.
For real $H_0$ and $V$, the roots come as $N_{\rm D}=\tfrac{1}{2}n(n-1)$ complex-conjugate pairs, so
\begin{equation}
D(\Lambda)\propto\prod_{\alpha=1}^{N_{\rm D}}(\Lambda-\Lambda_{\alpha})(\Lambda-\Lambda^*_{\alpha})
\,.
\label{Droot}
\end{equation}
The discriminant can be also expressed as \cite{Zir83}
\begin{eqnarray}
D(\Lambda)=\prod_{k<l}[E_l(\Lambda)-E_k(\Lambda)]^2=(-)^{N_{\rm D}}\prod_{k}D_k(\Lambda)
\label{Dfact}\,,\quad \\
D_k(\Lambda)=\prod_{l(\neq k)}[E_l(\Lambda)-E_k(\Lambda)]
\,,\quad
\label{Dk}
\end{eqnarray}
where we introduced a factorization by \lq\lq partial discriminants\rq\rq\ $D_k$.
The (unknown) functions $E_l(\Lambda)$ denote complex eigenvalues on individual Riemann sheets, which are pairwise connected by degeneracy points $\Lambda_{kl}$=$\Lambda_{lk}$ defined by $E_k(\Lambda_{kl})$=$E_l(\Lambda_{kl})$.
The set $\{\Lambda_{kl}\}$ with $k\!<\!l$ is equivalent to $\{\Lambda_{\alpha}\}$ with $\alpha$=1,\,$\dots,N_{\rm D}$.

If analyzing the $k^{\rm th}$state avoided crossings, only the degeneracies on the $k^{\rm th}$ Riemann sheet are relevant.
These are zeros of $D_k$.
However, with an increasing dimension $n$ the assignment of degeneracies to individual sheets becomes practically impossible.
Fortunately, there is a way how to indirectly measure the distribution of $k^{\rm th}$sheet degeneracies close to the {\em real\/} $\Lambda$ axis.
Squaring Eqs.~(\ref{Droot}) and (\ref{Dfact}) we find that $D_k^2$ is a polynomial of order $2(\nm1)$ with the roots $\Lambda_{kl}$ and $\Lambda^*_{kl}$.
On the real axis, $\Lambda=\lambda+i0$, one therefore has
\begin{equation}
D_k(\lambda)^2=
c_k\prod_{l(\neq k)}\underbrace{\left[(\lambda-\lambda_{kl})^2+\mu_{kl}^2\right]}_{R_{kl}(\lambda)^2}
\,,
\label{Dk2'}
\end{equation}
$c_k>0$, where $R_{kl}(\lambda)$ denotes the distance of the point $\Lambda_{kl}\equiv\lambda_{kl}+i\mu_{kl}$ from place $\lambda$.
Hence $D_k(\lambda)^2$ is sensitive to the proximity of the $k^{\rm th}$sheet degeneracies to the real axis and, at the same time, it is expressible through Eq.~(\ref{Dk}), using solely real eigenvalues of Hamiltonian (\ref{ham}).
By defining $U_k(\lambda)=-\tfrac{1}{2\Omega}\ln D_k(\lambda)^2$, where $\Omega=\nm1$ is a convenient scaling constant, we replace the product over degeneracies by a sum:
\begin{eqnarray}
U_k(\lambda)&=&-\tfrac{\ln c_k}{2\Omega}-\tfrac{1}{\Omega}\!\sum_{l(\neq k)}\ln R_{kl}(\lambda)
\label{Uk1}\\
&=&-\tfrac{1}{\Omega}\!\sum_{l(\neq k)}\ln|E_l(\lambda)-E_k(\lambda)|
\,.
\label{Uk2}
\end{eqnarray}
In the first line we identify a 2-dimensional Coulomb potential (scaled and shifted) caused by $\nm1$ point charges at positions $(\lambda_{kl},\mu_{kl})$ in the complex plane.
Therefore, the introduction of $U_k$ makes it possible to benefit from the basic intuition of planar electrostatics.
(Note that energy levels in the second line can also be thought as point charges.)

If the $k^{\rm th}$eigenstate exhibits a QPT at a certain critical point $\lambda^{\rm c}_k$ on the real axis, the degeneracies are expected to approach infinitely close to this point in the \Ninf\ limit \cite{Hei88,Cej05}.
Since the QPT depends most sensitively on the closest degeneracies, it is convenient to define the following additional quantities:
\begin{eqnarray}
F_k(\lambda)\!\!&=&\!\!-\tfrac{d}{d\lambda}U_k(\lambda)\,,\quad
C_k(\lambda)=\tfrac{d}{d\lambda}F_k(\lambda)
\,,\label{Ck}\\
Q_k(\lambda)\!\!&=&\!\!\lim_{\epsilon\to 0+}\int_{\lambda-\epsilon}^{\lambda+\epsilon}\!\!\!C_k(\lambda')d\lambda'
=\lim_{\epsilon\to 0+}[F_k]^{\lambda+\epsilon}_{\lambda-\epsilon}
\,.\quad
\label{Qk}
\end{eqnarray}
$F_k$ is nothing but a repulsive force felt by a \lq\lq trial charge\rq\rq\ on place $\lambda$ due to the charges in the complex plane.
As the charge distribution is symmetric against complex conjugation, the force is (anti)parallel with the real axis.
If the charges approach very close to $\lambda^{\rm c}_k$, the force will change its sign from $-$ to $+$ at this point and $C_k$, which represents a change rate of the force, will have a maximum there.
Sometimes the force may even jump at $\lambda^{\rm c}_k$, i.e. to flip its sign without actually crossing zero.
Such a place corresponds to asymptotic \lq\lq condensation\rq\rq\ of charges on the real axis (a locally 1-dimensional charge layer across ${\rm Im}\Lambda\!=\!0$ is formed) and can be detected by a nonzero value of $Q_k$ in Eq.~(\ref{Qk}).
In contrast, $Q_k$ is obviously zero as far as the force is continuous.

It turns out that the quantities defined in Eqs.~(\ref{Uk1})--(\ref{Qk}) are useful for the classification of the $k^{\rm th}$level QPTs.
From Eq.~(\ref{Uk2}) we get
\begin{equation}
F_k(\lambda)=\tfrac{1}{\Omega}\sum_{l(\neq k)}(-)^{\phi_{kl}}
\frac{\tfrac{d}{d\lambda}[E_l(\lambda)-E_k(\lambda)]}{|E_l(\lambda)-E_k(\lambda)|}
\,,
\label{Fk}
\end{equation}
where $\phi_{kl}\!=\!0$ for $E_l\!>\!E_k$ and $\phi_{kl}\!=\!1$ otherwise.
This is a continuous function of $\lambda$ unless some of the derivatives $\tfrac{d}{d\lambda}[E_l-E_k]$ are discontinuous.
Since the discontinuity (if any) appears in the \Ninf\ limit, when the spectrum becomes infinite, a \lq\lq macroscopic\rq\rq\ fraction of nonanalytic energy differences needs to sum up to produce a finite jump of $F_k$ at a certain point.
This is naturally satisfied if $\tfrac{d}{d\lambda}E_k$ itself has a jump, i.e. if the $k^{\rm th}$state exhibits a first-order QPT.
Therefore, $Q_k\!\neq\!0$ for a generic first-order transition while $Q_k\!=\!0$ for a continuous transition or for a crossover.
The latter two cases can be distinguished by the behavior of $C_k$, which shows an asymptotically singular peak at $\lambda^{\rm c}_k$ if there is a QPT of any type.

To illustrate these matters, let us give two examples.
In the first one we assume that the $k^{\rm th}$sheet degeneracies are all located on a line perpendicular to the the real $\Lambda$ axis at $\lambda\!=\!\lambda^{\rm c}_k$ and that they
become infinitely dense as \Ninf.
Introducing $\delta\equiv\lambda\!-\!\lambda^{\rm c}_k$ we obtain
\begin{equation}
C_k(\delta)=\int_0^{\infty}\!\!\rho_k(\mu)\,\frac{\mu^2-\delta^2}{(\mu^2+\delta^2)^2}\,d\mu
\,,
\label{Cklin}
\end{equation}
where $\rho_k(\mu)$ is an asymptotic linear density of degeneracies normalized to unity: $\rho_k(\mu)\!=\!\lim_{N\to\infty}\tfrac{1}{\Omega}\sum_l\delta(\mu\!-\!\mu_{kl})$.
Note that the normalization is ensured by the factor $1/\Omega$ which takes the reciprocal value of the number of degeneracies.
In any case, $C_k$ has a symmetric maximum at $\delta$=0 where all its odd derivatives (if they exist) vanish.
It becomes nonanalytic if the support of $\rho_k(\mu)$ has its infimum at $\mu$=0.
In particular, let us assume that close to this point the density can be approximated by $\rho_k(\mu)\sim\mu^p$, with $p\!\geq\!0$.
Then we can show that: $Q_k(0)$ is finite for $p\!=\!0$ and zero otherwise, $C_k(0)$ diverges for $p\!\in$(0,1] and is finite for $p\!>$1, $\tfrac{d^2}{d\delta^2}C_k(0)$ diverges for $p\!\in$(1,3] and is finite for $p\!>$3, $\tfrac{d^4}{d\delta^4}C_k(0)$ diverges for $p\!\in$(3,7] and is finite for $p\!>$7 etc.
Therefore, the line geometry of degeneracies induces a first-order QPT if $p$=0, i.e. $\lim_{\mu\to 0}\rho_k(\mu)\neq 0$.
In all the other cases the QPT is of a continuous type.
An increasing power $p$ shifts the $\delta$=0 nonanalyticity to higher and higher derivatives of $F_k$ (and $E_k$).
This sheds light on the asymptotic nature of level crossings in the first-order and continuous QPTs.

\begin{figure}
\epsfig{file=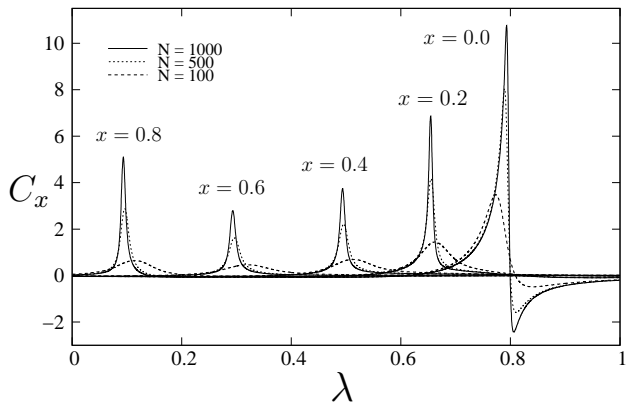,width=\linewidth}
\caption{\protect\small
$C_k=-\tfrac{d^2}{d\lambda^2}U_k$, see Eq.~(\ref{Uk2}), obtained from the $J$=$v$=0 spectrum of Hamiltonian (\ref{Hibm}) for some selected values of $k\!=\!xn$ and $N$. The peaks indicate the passage of a given level through an avoided crossing at $E$=0 \cite{Hei06}.
\label{fi2}}
\end{figure}

The second example shows the behavior of nonhermitian degeneracies in the interacting boson model of nuclear physics \cite{Iac87}.
It is one of the above-mentioned $sb$ bosonic models in which $b^{(L)}$ is identified with the $L\!=\!2$ boson called $d$.
The ground-state QPTs between spherical and deformed phases can be of both first and second order \cite{Die80}.
Here, we will focus on the [O(6)-U(5)]$\supset$O(5) transition \cite{Hei06}, which exhibits a second-order QPT for the ground-state and a chain of continuous QPTs for excited states \cite{Cej06,Cap07}.

The Hamiltonian per boson reads as
\begin{equation}
H(\lambda)=\lambda\,\tfrac{1}{N}n_d-(1-\lambda)\,\tfrac{1}{N^2}\!\!\!\sum_{m=-2}^{+2}\!(-)^m Q_m Q_{-m}
\,,
\label{Hibm}
\end{equation}
where $n_d$ stands for the $d$-boson number operator and $Q_m=d^{\dag}_m s+(-)^m s^{\dag} d_{-m}$.
Eq.~(\ref{Hibm}) has the form (\ref{ham}) with $\lambda\in[0,1]$, the value $\xi\!=\!1\!-\!\lambda$ rising the strength of interactions between $s$ and $d$ bosons.
The ground-state QPT is located at $\lambda_0^{\rm c}\!=\!\tfrac{4}{5}$, where the classical potential corresponding to Hamiltonian (\ref{Hibm}) changes between sombrero-like ($\lambda\!<\!\lambda_0^{\rm c}$) and quartic oscillator ($\lambda\!>\!\lambda_0^{\rm c}$) forms \cite{Die80}.
Excited-state QPTs take place for $\lambda\!<\!\lambda_0^{\rm c}$ at absolute energy $E$=0, where the sombrero potential has the central maximum \cite{Cej06}.
As individual levels cross the top of this maximum, wave functions become singular and level energies evolve with locally infinite curvatures \cite{Cap07}.
This is related to the crossing of the classical phase-space separatrix, which exists only in absence of the centrifugal barrier, i.e. for states with vanishing O(3) and O(5) relative \lq\lq angular momenta\rq\rq\ $J/J_{\rm max}$ and $v/v_{\rm max}$ \cite{Cej06}.
For finite $N$, the $J$=$v$=0 level dynamics shows a characteristic pattern of avoided crossings propagating through the spectrum at $E$=0 \cite{Hei06}.
The passage of a given level $k$ through the crossing yields a peak of $C_k(\lambda)$, as illustrated for some selected levels in Fig.~\ref{fi2}.
Here we replace integer $k$ by an excitation ratio $x=k/n\in[0,1]$.
The peak centroid moves leftwards linearly with increasing $x$ and sharpens with an increasing boson number $N$.

\begin{figure}
\epsfig{file=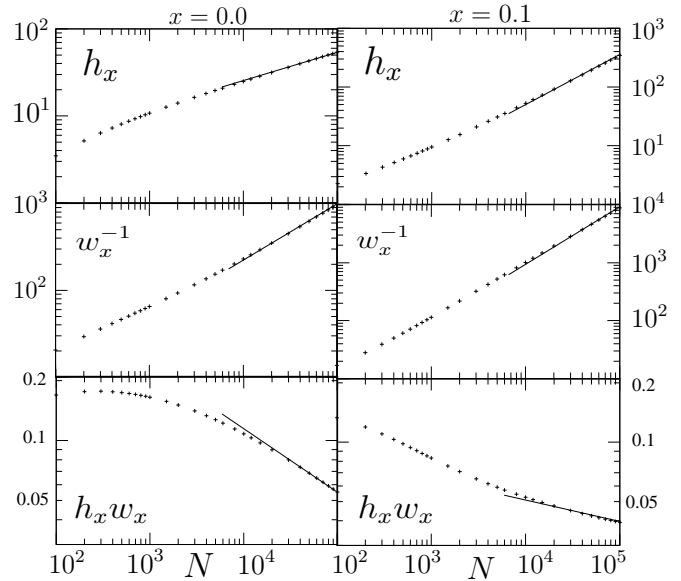,width=1.05\linewidth}
\caption{\protect\small
Large-$N$ evolution of the $x$=0 and $x$=0.1 peaks of $C_x(\lambda)$ for Hamiltonian (\ref{Hibm}).
Log-log plots from top to bottom show the peak height (relative units), inverse width at half maximum, and a product height\,$\times$\,width.
Lines indicate a power-law asymptotics for $N\!>\!10^4$. 
\label{fi3}}
\end{figure}

From semiclassical arguments it is known that the QPTs of Hamiltonian (\ref{Hibm}) are continuous, with discontinuous or infinite $\tfrac{d^2}{d\lambda^2}E_k$ for $x$=0 or $x\!>$0, respectively \cite{Cej06,Cap07}.
In finite-$N$ calculations, however, one can never verify whether $Q_x$ is zero at $\lambda_x^{\rm c}$ or not.
To address this problem, we determined the dependence $C_x(\lambda)$ for $x$=0 (the ground state) and $x$=0.1 (excited state in 10\,\% of the spectrum) for boson numbers up to $10^5$.
From these calculations we extracted the behavior of the peak height, $h_x$, and the peak width at half maximum, $w_x$.
Results are shown in Fig.~\ref{fi3}.
The peak height/width increases/decreases with $N$, both dependences being approximately of the power-law type for very large $N$.
The width decrease is faster than the height increase, so the product $h_x w_x$ decreases.
Since the product approximates the peak area, its \Ninf\ limit should coincide with $Q_x(\lambda_x^{\rm c})$.
We may therefore conclude that numerical data shown in Fig.~\ref{fi3} are compatible with the theory presented above, though the convergence to the asymptotic regime (presumably of a power-law type) is extremely slow.
Let us stress that the same type of behavior is expected in all $sb$ bosonic models in transition between their O(2$L$+2) and U(2$L$+1) dynamical symmetries \cite{Cap07}.

The above-described classification of QPTs in terms of nonhermitian degeneracies appears to be very similar to the classification of thermodynamic phase transitions in terms of zeros of the partition function ${\cal Z}$ in a complex-extended temperature plane \cite{Yan52,Bor00}.
This analogy has been proposed in Ref.~\cite{Cej05}, but it remained just a surmise.
Here, we can show that the two classifications are in fact identical.
Recall basic expression for the free energy ${\cal U}\!=\!-T\ln{\cal Z}$, specific heat ${\cal C}\!=\!-T\tfrac{\partial^2}{\partial T^2}{\cal U}$, and latent heat ${\cal Q}\!=\!\int{\cal C}\,dT$, where $T$ is temperature.
If neglecting unimportant prefactors, one sees that these quantities are in the same relations as functions $U_k$, $C_k$, and $Q_k$ from Eqs.~(\ref{Uk1})--(\ref{Qk}).
This link is valid if {\em formally\/} associating the partition function ${\cal Z}$ with an appropriate power of the partial discriminant $D_k$, hence also zeros of ${\cal Z}$ with the $k^{\rm th}$sheet degeneracies.
Therefore, the quantities introduced here purely on the basis of the Coulomb analogy for nonhermitian degeneracies have direct thermodynamic counterparts used in the description of standard phase transitions.
For instance, the criterion for the \lq\lq force discontinuity\rq\rq\ in the first-order QPT, $Q_k\!\neq$0, is just the familiar rule of nonzero latent heat.

In conclusion, we have presented a method to measure the distribution of nonhermitian degeneracies close to real values of the control parameter for Hamiltonians of the form (\ref{ham}).
It is based on electrostatic intuition, associating degeneracies in the complex plane with point charges, and makes it possible to separate degeneracies on Riemann sheets corresponding to different levels.
We formulated criteria for the first-order and continuous QPTs affecting an arbitrary level and tested them in the interacting boson model.
Finally, we explained the analogy between nonhermitian degeneracies and zeros of partition function.
These general results can be applied to nuclear, molecular, optical, and mesoscopic systems.

A discussion with U. G\"unther is gratefully acknowledged.
This work was supported by the Czech Science Foundation (202/06/0363), the Czech Ministry of Education (MSM\,0021620859), and the German Research Foundation (TSE 17/1/06).

\end{document}